\documentclass[11pt, a4paper]{article}
 
\usepackage[textwidth = 410pt]{geometry}
\usepackage{apacite}
\usepackage{etoolbox}
\renewenvironment{APACrefURL}[1][]{}{}
\AtBeginEnvironment{APACrefURL}{\renewcommand{\url}[1]{}}
\renewcommand{\doiprefix}{doi:~\kern-1pt}

\usepackage{xcolor}
\usepackage{microtype}
\usepackage{graphicx}
\usepackage[font=small]{caption}
\usepackage{lscape}
\usepackage{xurl}

\title{Recontextualized Knowledge and Narrative Coalitions on Telegram}

\author{
Tom Willaert$^{1, 2}$\\
$^{1}$Brussels School of Governance, Vrije Universiteit Brussel, Belgium\\
$^{2}$imec-SMIT, Vrije Universiteit Brussel, Belgium\\
\underline{tom.willaert@vub.be}
}

\date{}

\begin{document}

\maketitle

\centerline{\textcolor{red}{\textbf{Book chapter submitted to \textit{Advances in Sociolinguistics (Bloomsbury)}}}}

\begin{abstract}

A defining characteristic of conspiracy texts is that they negotiate power and identity by recontextualizing prior knowledge. This dynamic has been shown to intensify on social media, where knowledge sources can readily be integrated into antagonistic narratives through hyperlinks. The objective of the present chapter is to further our understanding of this dynamic by surfacing and examining 1) how online conspiracy narratives recontextualize prior knowledge by coupling it with heterogeneous antagonistic elements, and 2) how such recontextualizing narratives operate as connectors around which diverse actors might form narrative coalitions. To this end, the chapter offers an empirical analysis of links to prior knowledge in public messaging channels from the Pushshift Telegram dataset. Using transferable methods from the field of bibliometrics, we find that politically extreme Telegram channels engage with a variety of established knowledge sources, including scientific journals, scientific repositories and other sources associated with the system of scholarly communication. Channels engaging with shared knowledge sources thereby form narrative coalitions ranging from scientific and technological imaginaries to far-right extremist and antisemitic conspiracy theories. Our analysis of these coalitions reveals (i) linguistic, political, and thematic forces that shape conspiracy narratives, (ii) emerging ideological, epistemological and ontological positions associated with online conspiracism, and (iii) how references to shared knowledge contribute to the communicability of conspiracy narratives.    

\end{abstract}

\section{Introduction}

If textual genres can be analysed in terms of intertextual relations to other discourses \cite{briggs1992genre}, then the genre of conspiracy theory may well be defined by its recontextualizations of prior knowledge. By referencing academic papers, monographs, websites, and other knowledge sources, conspiracy texts appropriate the citation apparatus of ``conventional scholarship'' in order to lay claim to evidence-based observations about the world \cite[p. 7]{barkun2013culture}. Such references thus constitute a key instrument by which conspiracy texts negotiate power and identity, as they allow texts to assert authority over the established system of knowledge production, subsequently subverting it to lend credibility to antagonistic and other potentially harmful narratives \cite{milani2020nogo,rothermel2023role}. Previous research has demonstrated that this recontextualization of knowledge is intensified on social media \cite{tuters2023science}, where the use of hyperlinks (which may directly point to external sources) or hashtags (which may connect content with over-arching themes) affords the creation of structural couplings between heterogeneous forms of knowledge and discourse \cite{costas2021heterogeneous,mason2019intertextuality}.  

The objective of the present study is to further our understanding of how references to prior knowledge relate to the formation of narratives and narrative coalitions on social media. The chapter thus aims to surface and examine two structural functions of online conspiracy narratives, namely: 1) how conspiracy narratives recontextualize prior knowledge by coupling it with seemingly heterogenous elements, and 2) how such recontextualizing narratives operate as connectors around which diverse actors might form narrative coalitions. 

To this end, we conduct an empirical analysis of references to prior knowledge on Telegram, a messaging platform that has become strongly associated with the propagation of conspiracy theories and other forms of disinformation \cite{peeters2022telegram,willaert2022disinformation,simon2022linked}. This can be attributed to Telegram's loose stance on content moderation, which has been shown to attract extreme actors previously ``deplatformed'' from other social media \cite{rogers2020deplatforming}. Telegram offers such actors the possibility of creating one-way, public messaging channels through which they may broadcast content to large audiences of followers. As messages can be forwarded freely between channels, the platform removes much of the friction that might impede the circulation of conspiracy narratives \cite{willaert2023computational}. 

Building on these observations, the present chapter specifically traces the uses of hyperlinks to external knowledge in  messages from the multilingual Pushshift Telegram Dataset \cite{baumgartner2020pushshift}, which contains the contents of 28,000 Telegram channels ``snowballed'' from primarily English-language seed channels pertaining to right-wing extremist politics and cryptocurrencies, and which covers a period between 2015 and 2019. By zooming in on the object of the hyperlink, and specifically hyperlinks introduced in the textual body of messages in public Telegram channels, we aim to uncover the dynamics through which conspiracy narratives and narrative coalitions form on the platform. Our approach is thereby informed by two theoretical assumptions about the nature of conspiracy narratives and the communities in which they propagate. 

First, we subscribe to the notion that on a conceptual level, conspiracy narratives, just like political narratives more generally, may be characterized as sense-making devices \cite{bruner1991narrative} through which individuals and groups ``construct disparate facts in [their] worlds and weave them together cognitively'', enabling them to ``understand themselves as political beings'' \cite{patterson1998narrative}. Channelling Mark Fenster, we are thus primarily concerned here with how conspiracy narratives operate as ``organization[s] of data'' \cite[p. 120]{fenster2008conspiracy}, in which real-world knowledge and observations (as referenced through hyperlinks) become associated with seemingly heterogeneous concepts to form coherent explanations for the world's workings. Drawing on insights from structuralist narratology, conspiracy narratives might thus be analysed as devices that organize representations of people, places and other real-world observations according to underlying plot structures. Archetypical examples include Freytag's narrative arc \cite{freytag1895technique}, Propp's morphology of the folktale \cite{propp1968morphology}, or Campbell's hero's journey \cite{campbell2008hero}. Likewise, narratives may organize their constituent parts according to ``actantial'' models such as Greimas' actantial narrative schema \cite{greimas1987actants,herman2005actant}, which comprises structural relations between entities (e.g. concepts, characters). This actantial perspective has proven productive for the analysis of conspiracy narratives in social media environments, as it lends itself to operationalizations in terms of networks and network analysis \cite{tangherlini2020automated}. Research following this approach has thereby demonstrated that conspiracy narratives in participatory web environments are marked by dynamics of intensified ``narrative convergence'' \cite{tuters2022deep,willaert2023computational}, in which disparate concepts are rapidly linked together into over-arching narratives - a dynamic that elsewhere has been referred to as a ``conspiracy singularity'' \cite{klein2020great,merlan2020conspiracy}. 

Secondly, in addition to organizing heterogeneous concepts on a structural level, conspiracy narratives have been demonstrated to contribute to the formation and organization of communities \cite{bessi2015science}. This observation recalls Maarten Hajer's concept of ``discourse coalitions'', wherein ``different actors from various backgrounds form specific coalitions around specific story lines. Story lines are the medium through which actors try to impose their view of reality on others, suggest certain social positions and practices, and criticize alternative social arrangements.'' \cite[p. 47]{hajer2002discourse}. Following this description, the objective of the present chapter is to identify what we will refer to as the ``narrative coalitions'' that form around shared emplotments of prior knowledge in social media. The chapter thus (simultaneously) addresses the following empirical research questions: 

\begin{enumerate}
\item Which knowledge sources are referenced in conspiratorial Telegram channels and which heterogeneous concepts are they associated with to form narratives?
\item What are the latent narrative coalitions that emerge when we examine associations between Telegram channels based on such shared references to prior knowledge? 
\end{enumerate}

As recontextualized knowledge constitutes a core element of conspiracy theories, we expect this empirical tracing of knowledge sources on Telegram to yield more general insights into the thematic, linguistic, and political forces that shape online conspiracy narratives.

\section{Methodology}

\subsection{Bibliographic coupling analysis}

In order to mine traces of the (ab)uses of knowledge sources from large collections of Telegram messages, we turn to transferable methods form the quantitative field of bibliometrics. Pioneered by scholars Eugene Garfield and Derek John de Solla Price in the 1960s, bibliometrics is primarily concerned with ``gauging the perception of academic publications'' on the basis of citations \cite[p. 2]{ball2021introduction}. The central principle underpinning its wide range of methods and metrics, is that publications that are cited frequently might be considered important to a field of research, whereas those cited less frequently might be considered less relevant \cite[p. 2]{ball2021introduction}. Today, the field has become negatively associated with academic power structures and the evaluation of scholars or institutions based on the quantification of scientific output \cite[p.30]{pendlebury2021eugene}. Yet we adhere here to a more descriptive interpretation of bibliometrics, which is closely aligned with its original purposes, and in which its methods are primarily used to map ``knowledge flows'', the ``diffusion of ideas'', and the underlying ``intellectual structures'' that mark a given field of research \cite[p.2]{zhao2015analysis}. Rather than investigating references in traditional scholarly publications such as books or journal articles, we focus on hyperlinks referencing knowledge sources in conspiratorial Telegram channels. This transfer of methods hinges on the aforementioned observation that conspiratorial texts and discourse have been argued to mimic the scholarly apparatus of references and citations \cite[p. 7]{barkun2013culture}. 

\begin{figure}[!htbp]
  \centering
  \includegraphics[width=0.5\textwidth]{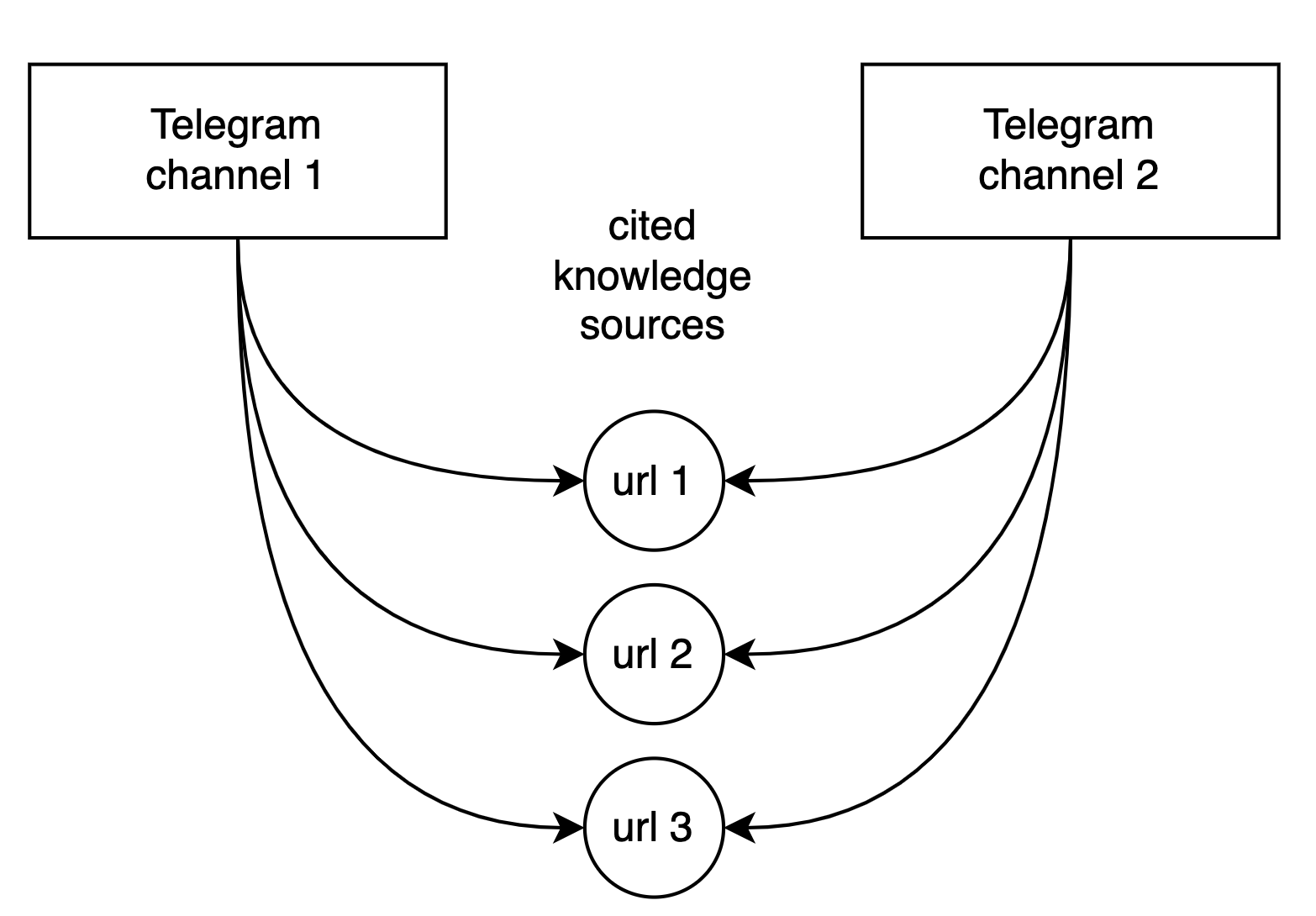}
  \caption{Illustration of bibliographic coupling of two Telegram Channels. Both Telegram channel 1 and Telegram channel 2 contain at least on reference to url1, url2 and url3. As such it can be argued that both channels are connected with a bibliographic coupling frequency of 3.}
  \label{fig:methodology_BCF}
\end{figure}

At the center of our investigation is a bibliographic coupling frequency (BCF) analysis, which quantifies the strength of association between two papers based on the number of references they share. It is thereby assumed that papers sharing references to the same knowledge sources might also be intellectually similar to each other \cite[p. 541]{garfield1968world}. This reasoning extends to the present situation in the sense that we may argue that Telegram channels sharing links to similar knowledge sources can be considered to belong to the same ``narrative coalition'' (see Figure \ref{fig:methodology_BCF}).

As demonstrated by previous work, a BCF approach readily lends itself to operationalizations in the context of large collections of data mined from social media \cite{yang2022map}. Our data processing pipeline is thus geared towards preprocessing and visualizing our collection of Telegram messages in ways that directly speak to our research questions: first, it allows us to produce an overview of the different types of knowledge sources referenced in the Telegram channels, and secondly, it offers a network view of the communities (narrative coalitions) among channels that can be identified by means of a BCF analysis.

\subsection{Data processing pipeline}

\begin{figure}[!htbp]
  \centering
  \includegraphics[width=0.5\textwidth]{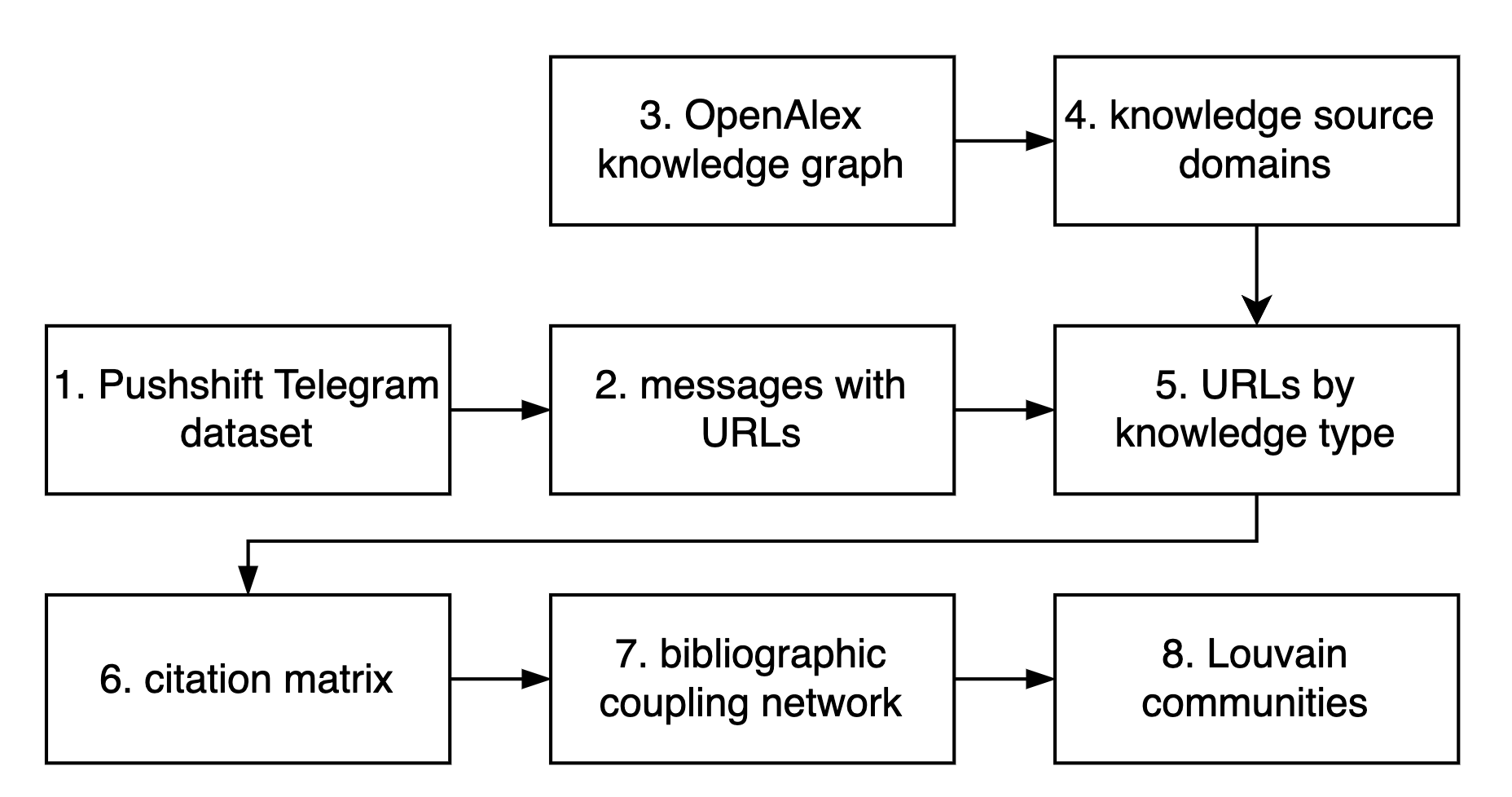}
  \caption{Overview of the main steps in our processing pipeline. }
  \label{fig:methodology_pipeline_diagram}
\end{figure}

The main steps in our pipeline are illustrated in Figure \ref{fig:methodology_pipeline_diagram}. In a first step, we iterate over each message included in the Pushshift Telegram Dataset \cite{baumgartner2020pushshift}, which contains 317 million Telegram messages posted by 2.2 million users in 28,000 channels (covering both public broadcasting channels and chat groups) between 2015 and 2019. From each message, we extract the URLs that are included in the message text. We thus retain a dataset of 12,938,085 messages that contain at least one URL. If a URL is a shortened url, we first expand it using the python urlExpander library \cite{leon2018urlexpander}.
We then filter out the second-level domain from each URL, that is: the domain directly below the top-level domain, which might, among other things, refer to a country code (e.g. ``.be'') or an organization (e.g. ``.org''). We subsequently normalize the URL strings for further classification. The URL \url{https://www.journals.elsevier.com/fuel-and-energy-abstracts} is thus converted to ``elsevier.com''.\footnote{We make an exception here for the  Google subdomains ``books.google.com'', ``sites.google.com'', ``scholar.google.com'', ``docs.google.com'', ``drive.google.com'', which are retained in full. Other Google subdomains (e.g. ``play.google.com'') are reduced to ``google.com''.}

Next, we classify each simplified URL according to whether it refers to an established knowledge source or not. We approach this classification inductively by referencing each simplified URL in the Telegram dataset against the list of sources included in the OpenAlex knowledge graph \cite{priem2022openalex}. OpenAlex is a citation database that presents itself as a ``fully open catalog of the global research system'' that be queried via an API (\url{https://docs.openalex.org/}). The knowledge base offers an open alternative to paywalled knowledge bases such as Scopus or Web of Science, and indexes 249,000 academic sources, including ``journals, conferences, preprint repositories, and institutional repositories'' (\url{https://docs.openalex.org/api-entities/sources}). These sources are identified and aggregated from other projects including the Directory of Open Access Journals (DOAJ), Pubmed, the Internet Archive and institutional repositories (\url{https://docs.openalex.org/additional-help/faq}). In order to construct a reference corpus for classifying the URLs retrieved from the Telegram dataset,  we compile a list of all source URLs for works contained in the knowledge graph (\url{https://docs.openalex.org/api-entities/sources}). These URLs are then filtered and cleaned following the same procedure as the URLs extracted from the Telegram dataset, thus retaining a list of 18,494 domains of OpenAlex sources.

In a next step, we encode each of the retrieved URLs from the Telegram dataset according to whether it occurs in the OpenAlex knowledge graph or not, focusing our analysis on Telegram's public broadcasting channels. This results in an enriched dataset that we aggregate on channel level, retaining for each channel the counts of the URLs per type (OpenAlex knowledge object or not), as well as metadata including the channel title and channel description. We then proceed to submit the dataset of channels to a bibliographic coupling analysis, in which we quantify associations between channels on the basis of the number of references to knowledge sources they share between them. To this end, we follow the method outlined in \citeauthor{zhao2015analysis} \citeyear{zhao2015analysis}, pp. 31-39.

\begin{figure}[!htbp]
  \centering
  \includegraphics[width=0.5\textwidth]{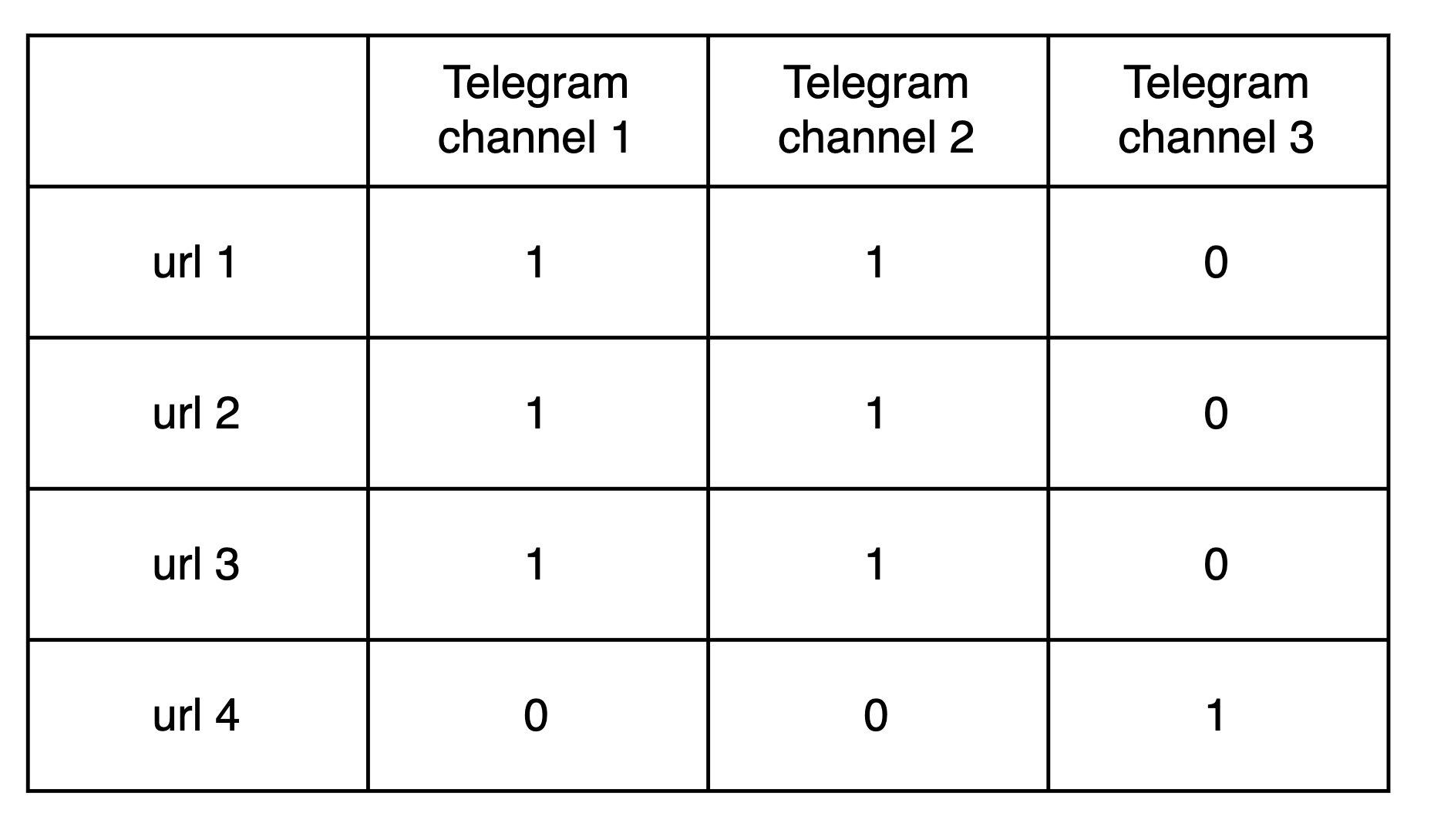}
  \caption{Illustration of a simple citation matrix. Each column represents a Telegram channel, rows represent sources. If a channel refers to a source, we set the value to 1, otherwise we set the value to 0.}
  \label{fig:methodology_citation_matrix}
\end{figure}

First, we construct a citation matrix, in which we keep track of the relationship between channels (columns) and sources (rows) (see Figure \ref{fig:methodology_citation_matrix}). If a source is referenced by a channel (i.e. if there is at least one message in the channel that cites it), we set the corresponding row value for that source to 1, if not, we set it to 0. In this step, we reduce the dataset by excluding the five sources that are cited by the highest number of channels (i.e. those with the highest row sum in our citation matrix). We omit these channels because they connect large numbers of channels in our bibliographic coupling graph graph, and as such they would obscure the more fine-grained intellectual structures of the network. We provide an overview of these highly frequent sources in the Findings section.

Conceptually, the citation matrix, which maps connections between channel names and knowledge sources, provides initial insights into structural connections between knowledge sources and channels. Channel names can thereby be considered important indicators of the ``heterogeneous'' narratives that these sources are enveloped in.

Based on the citation matrix, we proceed to calculate the bibliographic coupling frequency between each pair of Telegram channels in our dataset. Following the example in Figure \ref{fig:methodology_citation_matrix}, we can do this by calculating the scalar product of their corresponding vectors (columns) \cite[p. 39]{zhao2015analysis}. For example, the BCF of Telegram channel 1 and Telegram channel 2 is calculated as V(1,1,1,0) x V(1,1,1,0) = 3. The BCF between Telegram channel 1 and Telegram channel 3 is calculated as V(1,1,1,0) x V(0,0,0,1) = 0.  

We then construct a graph representation of connections between channels based on the bibliographic coupling frequency. In this graph, which we will refer to as a bibliographic coupling network, each node represents a channel, and each edge between nodes represents the bibliographic coupling frequency (BCF) between those nodes. We then use methods from network analysis to analyse the structural properties of the bibliographic coupling network. Specifically, we use the Louvain community detection algorithm to find non-overlapping groupings of nodes in the graph \cite{blondel2008fast}. Conceptually, this approach yields insights into possible narrative coalitions among channels, that is: communities of conspiratorial or otherwise antagonistic channels that engage with similar knowledge sources.
\section{Findings}

\subsection{Heterogenous couplings}

In the Pushshift Telegram dataset, we find 24,198 public broadcasting channels that contain URLs. Out of these, we extract the top 10,000 channels by the total number of references to sources that also occur in the OpenAlex knowledge graph. This results in a citation matrix with 10,000 channels and 77,747 different cited sources (including OpenAlex and non-OpenAlex sources).

\begin{figure}[!htbp]
  \centering
  \includegraphics[width=0.9\textwidth]{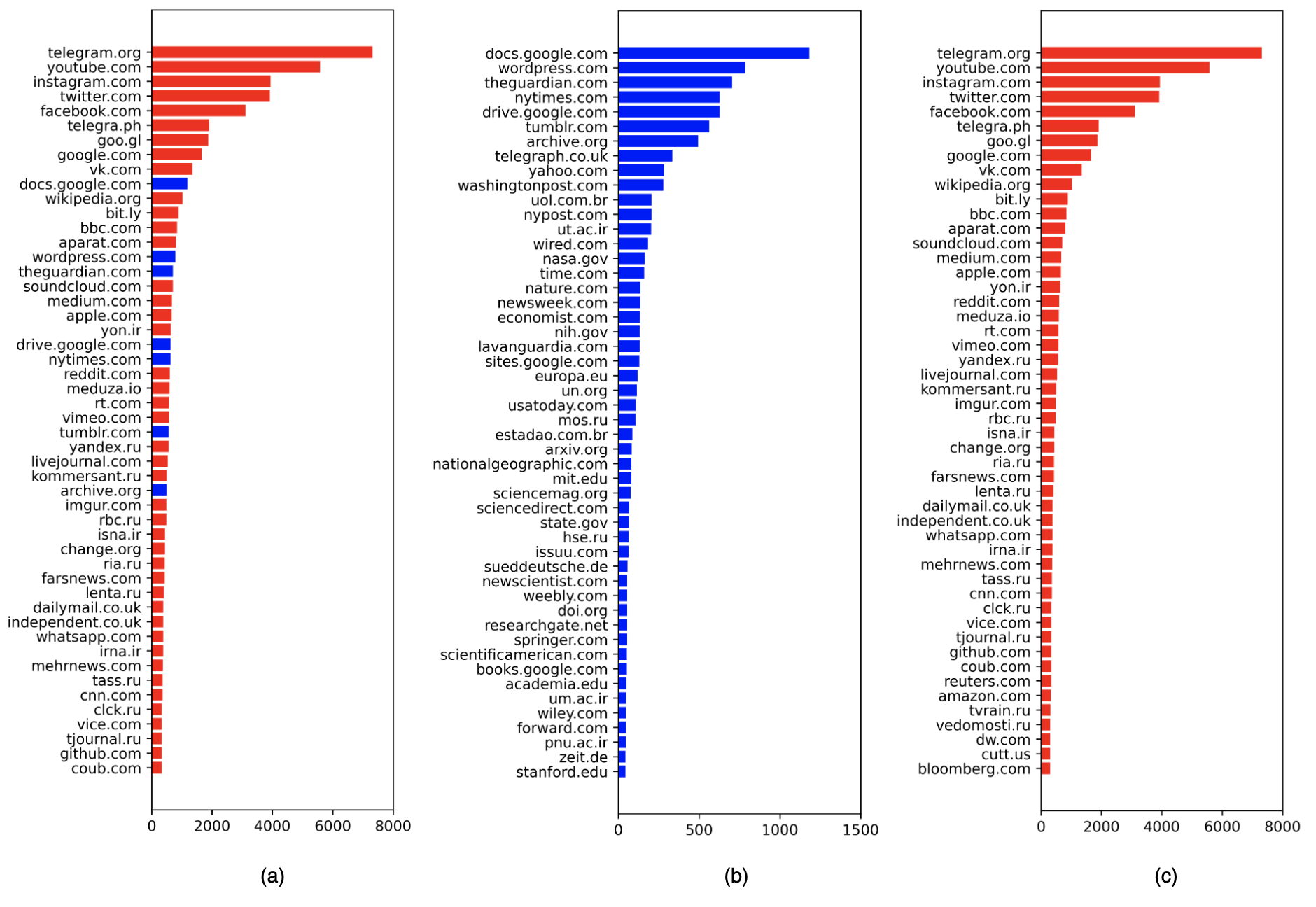}
  \caption{Overview of 50 most frequent domains in top 10,000 channels by number of links to OpenAlex sources, segmented by: combined source types (a), OpenAlex sources (b), non-OpenAlex sources (c)}
  \label{fig:findings_domain_frequencies}
\end{figure}

As follows from Figure \ref{fig:findings_domain_frequencies}(a), we find that the Telegram channels in our dataset refer to a heterogeneous set of websites. 7 of the 50 URLs that are cited by the highest number channels thereby also occur in the OpenAlex knowledge graph. As shown in Figure \ref{fig:findings_domain_frequencies}(c), the list of non-OpenAlex sources cited by most channels first and foremostly contains what elsewhere has been referred to as ``Very Large Online Platforms and Search Engines'' (VLOPS) \cite{ec2024dsa}. This includes references to Youtube, Instagram, Twitter (presently named ``X''), Facebook, and Wikipedia. Of the non-VLOPS, Telegram and its text and editor and publisher Telegraph figure prominently in the list (``telegram.org'', ``telegra.ph''). Additionally, the list also contains alternatives to these established platforms, such as the Russian social networking site VKontake (``vk.com''), the Iranian video sharing platform Aparat (``aparat.com''), and the search engine Yandex (``yandex.ru''). Likewise the list contains references to established news websites (``bbc.com''), and state-owned counterparts, such as Russia Today (``rt.com'').\footnote{From our overview of most cited sources, it follows that both the OpenAlex and non-OpenAlex lists contain references to news outlets. However, it seems to be the case that OpenAlex prioritizes actual publications (e.g. ``time.com'') over general news websites (e.g. ``bbc.com''). This could be explained on the grounds of the citability of the former: academic publications might be more likely to reference the contents of specialized news publications.} On a methodological note, it should be pointed out that the domains ``goog.le'' and ``bit.ly'' refer to shortend URLs that could no longer be expanded. This might be attributed to changes in the target domain (removal, migration to another domain) that might have happened between the collection of this dataset (with the most recent posts dating from 2019) and the moment of URL expansion (2024).

As shown in Figure \ref{fig:findings_domain_frequencies}(b), the OpenAlex sources cited by the highest numbers of channels comprise a variety of sources that (by virtue of their inclusion in the knowledge graph), have at some point become associated with the established ``apparatus'' of scholarly communication. This includes file sharing and archiving websites (``docs.google.com'', ``drive.google.com'', ``archive.org''), and established news sources (``theguardian.com'', ``nytimes.com'', ``washingtonpost.com''). We also find sources that are more central to scholarship and scholarly communication, including references to scientific journals and domains of scientific publishers (``nature.com'', ``sciencemag.org'', ``springer.com'', ``wiley.com'', ``sciencedirect.com''), scientific repositories (``arxiv.org''), as well as digital object identifiers (DOIs) typically associated with scientific publications (``doi.org''). Further, we find references to domains of higher education institutions (``stanford.edu'', ``mit.edu'') and governmental organizations (``nih.org'', ``nasa.gov''). Finally, we also find references to scientific magazines (``scientificamerican.com'') and academic social networking websites (``researchgate.net'').

From this overview, it follows that Telegram channels snowballed from a seed of English-language, right-wing extremist channels actively engage with a variety of sources. One trend that emerges when we examine the non-OpenAlex sources, is that the Telegram channels in the dataset interact quite prominently with links referring to sources either based in or focused on Russia. Some of these (viz. Russia Today) are, at the moment of writing, banned in the European Union on the grounds of spreading disinformation \cite{chee2022russia}. While it might not come as a surprise that such sources figure in a dataset that has been characterized as a key resource for studying computer-mediated disinformation campaigns \cite[p. 6]{baumgartner2020pushshift}, their presence does foreground the rather heterogeneous nature of those links that refer to more established knowledge sources. How exactly do these sources, which are part of the OpenAlex knowledge graph, connect to the overall antagonistic nature of our corpus? In order to examine the exact nature of these ``heterogeneous couplings'' between prior knowledge and elements of conspiratorial narratives, we proceed to identify the narrative coalitions among channels that form around these sources.

\subsection{Narrative coalitions}

\begin{figure}[!htbp]
  \centering
  \includegraphics[width=\textwidth]{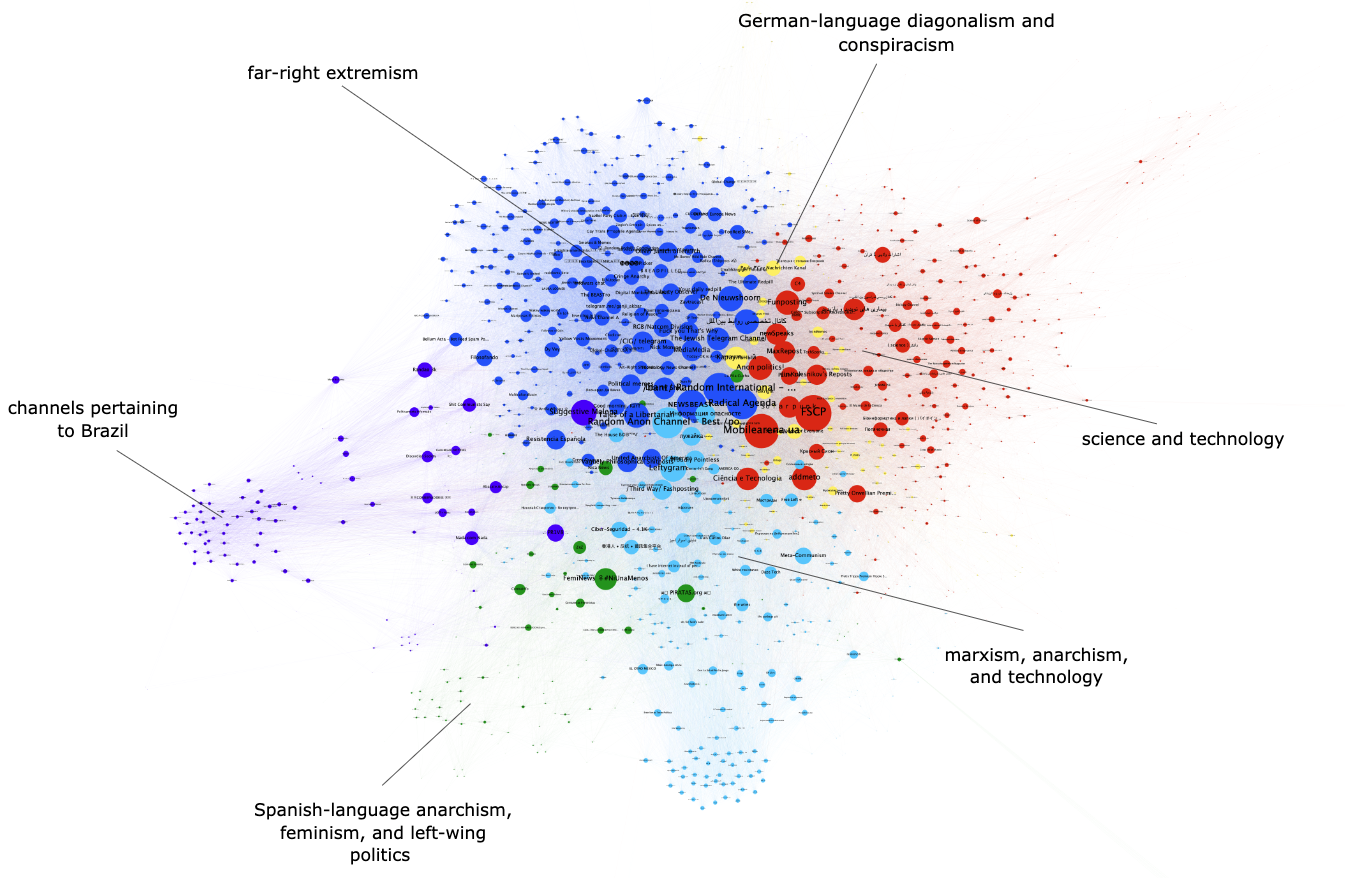} 
  \caption{Visualization of bibliographic coupling network of Telegram channels. Nodes represent channels, weighted edges the bibliographic coupling frequency (BCF) between two channels based on OpenAlex sources cited by both channels. Only edges representing a BCF higher than or equal to 2 are retained. Nodes are coloured according to Louvain communities (narrative coalitions). Node positioning: ForceAtlas2 algorithm, Node size: node degree.} 
  \label{fig:findings_BCF_graph}
\end{figure}

We analyse narrative coalitions based on a bibliographic coupling analysis of channels referencing sources included in the OpenAlex knowledge graph. To this end, we first filter our citation matrix by removing all non-OpenAlex sources, and then remove the five OpenAlex sources that are cited by the highest number of channels, notably: ``docs.google.com'' (cited by 1182 channels), ``wordpress.com'' (cited by 785 channels), ``theguardian.com'' (cited by 704 channels), ``nytimes'' (cited by 626) channels, and drive.google.come (cited by 626 channels). Based on this filtered citation matrix, we construct a bibliographic coupling matrix of 10,000 channels; (i.e. it contains 10.000 rows and 10.000 columns). We then construct a graph from the bibliographic coupling matrix, where nodes represent channels, and weighted edges the BCF between two channels. We only retain those edges that represent a bibliographic coupling frequency of at least 2. This yields a bibliographic coupling graph of 1382 nodes and 56,338 edges. We then inductively identify narrative coalitions using the Louvain Community detection algorithm. Each node in the graph is thereby coloured by the community to which it pertains. The resulting bibliographic coupling network is shown in Figure \ref{fig:findings_BCF_graph}.

As follows from this figure, we can identify 6 main communities in our dataset, which we will refer to as narrative coalitions. We thereby disregard three smaller communities, each of which contains less than 1\% of the channels in the dataset. A general visual inspection of the graph reveals three forces that shape the narrative alliances that can be observed. First, coalitions are marked by shared intellectual or thematic orientations, which may or may not reflect real-world events. Second, narrative coalitions are marked by a shared regional or linguistic dimension. Third, narrative coalitions can be distinguished from each other on the basis of their political orientation.

While there is a rich diversity of channels in each of these coalitions, whereby politically or thematically opposed channels may contend and connect over the same sources, the following sections provide a broad characterization and analysis of each of these coalitions based on visual inspections of the graph and the associations evoked by the names of the channels that make up the coalitions. As will follow, many of these channel names contain vernacular elements that can be considered indicative of subcultural communities and their idiosyncratic narratives \cite{peeters2021vernacular}. Our investigation thus reveals narrative coalitions pertaining to (i) scientific and technological imaginaries, (ii) far-right extremism and antisemitic conspiracy theories, (iii) anarchist and Marxist discourse, (iv) anti-vaccination conspiracy theories, (v) feminist and anti-feminist discourse, (vi) and national politics. 

\subsubsection{Science and technology} 

The largest narrative coalition in our dataset is broadly concerned with the fields of science and technology. It thus deals explicitly with questions of (scientific) knowledge, and shapes a series of idiosyncratic ontological and epistemological positions, some of which might be situated at the fringes of scientific inquiry.

This narrative coalition is primarily concerned with fleshing out ontological and aesthetic positions that imagine humanity's future position in light of technological advances. Channels thus recontextualize knowledge sources to shape concepts of transhumanism (``transhumanism\_in\_our\_hearts''), cyberpunk (``Subhuman Cyberpunk'') and solarpunk (``solarpunk'', ``lunarpunk''), as well as futurism (``Futurismo''), Orwellian dystopianism (``Pretty Orwellian Premium Morbid Reality'') and dataism (``data scientology'').

On an epistemological level, the coalition is marked by traces of skepticism (``skeptic salmon'') and rationalism (``the Indian Rationalist''). It also contains explicit references to alternative forms of knowledge and modalities for accessing knowledge, ranging from the neutral ``copyleft'' to refer to legal frameworks for open access to knowledge, to more marked forms of knowledge, such as ``pagan'' knowledge (``Pagan Knowledge'', ``The Viking Programmer''). Likewise, the narrative coalition contains references to the ``redpilling'' meme (``Redpill Dump''). This right-wing metaphor finds its origins in the 1999 science fiction movie \textit{The Matrix}, and refers to a process in which ``taking the red pill'' leads to an ``awakening'' or an ``epiphany about the rightness of white nationalism'' \cite{hagen2020reactionary}. 

The latter resonates with other overtly racist and subcultural vernacular channels in this coalition, as illustrated by channel names such as ``Racially charged memes and posts'', and ``Anon Politics'' (which recalls elements of the anonymous messaging board 4chan). Moreover, this coalition engages explicitly with instances of the Qanon conspiracy theory (``Qlobal-change Italia''). Closely associated with the ``pizzagate'' conspiracy theory,  QAnon revolves around the false narrative that established, wealthy elites form Satanic cabals and child molesting rings that are conspiring against Donald Trump \cite{tuters2022deep}. 

Interwoven with the abovementioned channels are channels that engage with a variety of scientific disciplines and subjects, including biology (``Science of Biology'', ``astrobiology''), mathematics (``Mathematics Association''), engineering (``Mechanical Engineering ( ut\_redc)''), robotics (``Robotrends.ru''), neuroscience (``neuroscienceru'', ``The Brains''), pharmaceutics (``Small Pharma'', ``Development of Pharmacy Channel''), psychology (``International Scientific Group on Positive Psychology''), the humanities (``philosophy bookshelf'', ``english book house'', ``Linguistics Library'', ``translation studies'', ``ecolinguistics''), spiritualism (``Spiritual Science Channel''), theology (``theology bites''), and atheism and agnosticism (``Atheists \& Agnostics''). The narrative coalition likewise contains marked stances on health and diet (``case against veganism''), and ``based'' reflections on education (``based education hub''). Another right-wing slang term, the word ``based'' has become associated with varying meanings revolving around the concepts of ``not caring about what other people think'' and ``being true to oneself'' \cite{hagen2023based}. 

While an in-depth discussion of the complex semantic and vernacular dimensions of this coalition, as well as the alternative knowledge it propagates are beyond the scope of this chapter, our overview clearly illustrates how a narrative coalition pertaining to scientific imaginaries can readily be marked by extreme political and ideological positions. The latter is reflected even more so in our second coalition.

\subsubsection{Far-right extremism} 

The second major narrative coalition in the dataset consists of channels explicitly associated with right-wing extremism, memes and elements pertaining to the ``deep vernacular web'' \cite{dezeeuw2020teh}. Of all coalitions, this one engages most overtly with antisemitic conspiracy theories. 

Interconnected channels in the dataset propagate the aforementioned ``Pizzagate'' conspiracy and the antisemitic conspiracy theories of Jewish cabals and ritualistic murder (e.g. ``Pizzagate/Ritualistic Jewish Cabals'', ``Jewish Ritual Murder Abortion Satanism''). It contains explicit traces of holocaust denial (``Shoahoax'') and antisemitic rhetoric (``Judenpresse Monitor/Archive'', ``K i K E s C e N t R a L''), the term ``kikes'' being a disparaging and offensive term to refer to Jews. Illustrative of an overall dynamic of contestation over knowledge, we find that these channels populate the same coalition as channels associated with news outlets of the Jewish community (e.g. `Jewish News TM''). 

Further conspiracy theory channels in this coalition include overtly transphobic conspiracy theories (``Gay Trans P**ophile Agenda''), as well as the ``white genocide'' conspiracy theory (``White Genocide Immigration Anti-White Agenda''), a white supremacist conspiracy theory that propagates unfounded beliefs of hidden plans geared towards the extinction of whites. The coalition furthermore contains channels that propagate conspiracy theories surrounding the ``New World Order'' (NWO) and the Illuminati (``Anti-nwo'',``Antiilluminaten TV''). As in the previous coalition, channels also explicitly refer to the QAnon Conspiracy theory and its ``where we go one we go all'' slogan (``Qlobal-change'', ``Brexiteers MBGA WWG1WGA''). These channels co-occur with channels marked by explicit evocations of Nazism and fascism (e.g. ``Fascist Library'', ``FacistBook News \& Media'', ``THE CATHOLIC WIGNAT'', ``NazBol Party Club'') as well as radical politics (e.g. ``Radical Agenda''). The coalition also includes channels pertaining to right-wing and conspiracist organisations (``Infowars chat'', ``ProudBoys'').  These channels are interconnected with channels evoking elements of the ``deep vernacular web'', typically associated with the language that characterizes platforms such as 4chan (e.g. ``CUCKISTAN'', ``Chadistan'', ``Dank Memes Gang'', 
``4Chan /POL/HIS/INT''), as well as the practice of shitposting (``Radical Shitposts'', ``Vaguely Philosophical Shitposts''). Just like the previous coalition, this cluster also contains explicit references to the concept of ``redpilling'' (e.g. ``Random Redpills Compilation'', ``The Ultimate Redpill'', ``Your daily redpill'', ``B R E A D P I L L E D''). If anything, this cluster of channels thus further emphasizes the extreme nature of the discourse and narratives that subsume prior knowledge on Telegram.

\subsubsection{Marxism, anarchism, and technology} 

The third narrative coalition that forms around established sources of knowledge is comprised of channels associated with left-wing politics, Marxism and anarchism, in particular as these relate to technology. Channels in the coalition thus refer to left-wing politics (e.g. ``Free Left'', ``Center-left Gang AMERICA GOOD edition'', ``LeftyGram''), as well as communism (``Meta-Communism'', ``National Bolshevik Ideology'') and anarchism (``ANARCHObois'', ``Anarquismo (canal)'', ``MemoriaAnarquista'', ``Anarcho-sheet''). 
In the same narrative alliance, we find channels dealing with digital technologies (e.g. ``e m a c s'', ``Linuxgram'', ``Stuff and Docs'', ``DOFH - DevOps from the hell'', 
``Technical Writing 101''), some explicitly opposing technology (``Neo-Luddite's Cabin''). Interconnected with these Marxist, anarchist and technology-oriented channels are channels pertaining to specific fandoms, such as furry culture, Taylor Swift, and the Teenage Mutant Ninja Turtles (``Taylor Swift Updates'', ``Furry Senate'', ``AltFurry Meme Factory'', ``TMNTFans''). This cluster likewise promotes a number of idiosyncratic positions (e.g. ``Pretty Trippy Premium Hippie Shit'', ``Macchiagram'').

This cluster speaks to the narrative convergence between political discourse, technological imaginaries, and popular culture. As will follow, the latter in particular contributes to what we may refer to as the ``communicability'' of conspiracy theory, disinformation and other antagonistic narratives.

\subsubsection{German-language diagonalism and conspiracism} 

The fourth narrative coalition in our dataset is a more heterogeneous coalition, situated at the intersections of previously discussed coalitions. This hetorogeneity is exemplified by channel names contriving concepts such as ``Black Gay Space Communism'', which recalls the ``Fully Automated Luxury Gay Space Communism'' meme ironically presenting a communist utopia. The most distinguishing feature of this coalition however is a cluster of German-language channels. These pertain to diagonalism or the ``querdenken'' movement (``NRW stellt sich quer Info''), as well as conspiracism (``Die Wahrheit und nur die Wahrheit. Was die offiziellen Medien nie zeigen.'',   ``Impfkritik'', ``Unabhängiger Infokanal'') and patriottic resistance (``Unabhängiger Infokanal''). This cluster of channels is clearly held together by an interaction with shared knowledge sources that cater specifically to the German context, which highlights a prominent regional and linguistic dimension of this type of online conspiracy discourse. 

\subsubsection{Spanish-language feminism, anarchism, left-wing politics} 

The fifth narrative coalition that forms around established knowledge sources is marked by Spanish-language channels pertaining to feminism, anarchism, and left-wing politics. This narrative coalition thus contains references to far-left political parties (``Partido Pirata de Chile'', ``Partido Popular'', ``Secretaría Rural PODEMOS'') and labour movements (``Movimiento laboral sindical''). The coalition also contains references to anarchism (``AnarcoCanal''), guerrilla warfare (``Guerrilla''), political counter information (``Contrainfo''), and environmentalism and sustainability (``BlogSOSTENIBLE - Ecología y más''). Strikingly, the coalition contains a cluster of channels referring to feminism (``FemiNews'', ``Comuneras Feministas'', ``CANAL CÍRCULOS FEMINISTAS PODEMITAS'', ``Economía feminita''). As in the case of the aforementioned coalitions, we again find traces of interactions with technology and cybersecurity ( ``Sombrero Blanco - Ciberseguridad''). Another recurring dynamic that can again be observed, is that the same sources are contended over by channels with diverging orientations. In the cluster dealing with feminism, we for instance find examples of channels taking up anti-feminist positions (e.g. ``El Feminismo es cáncer'').

\subsubsection{Channels pertaining to Brazil} 

The final narrative coalition that we identify in our dataset primarily consists of channels with titles in Spanish and Portuguese, many of which pertain to the context of Brazil. Within this thematically diverse cluster, we see tensions play out that also mark the dataset at large, notably opposing channels laying claim to the same sources of knowledge. As such, this heterogeneous cluster at the same time contains references to conservative politics (``CONSERVADORES'', ``Canal Conservador'') as well as LGBTQ+ channels (``LGBT+ News''), along with ``politically incorrect'' channels (``Politicamente incorreto''). As was the case for some of the aforementioned channels, these clusters are interlaced with references to technology and popular culture, such as references to the movie \textit{Fantastic Beasts and Where to Find them} (``Animais fantásticos e onde habitam'').

\section{Discussion}

In light of our twofold research question, we find that politically extreme channels in the Pushshift dataset engage with a variety of established knowledge sources, including scientific journals, scientific repositories and other sources associated with the system of scholarly communication. Channels engaging with shared knowledge sources thereby form narrative coalitions on scientific and technological imaginaries, far-right extremism and antisemitic conspiracy theories, Marxist and anarchist discourse, anti-vaccination conspiracy theories (in German), left-wing political discourse, and regional politics (pertaining to Brazil). 

These empirical observations about the use and abuse of knowledge contribute to our understanding of conspiracy texts on social media in three ways. First, our network visualizations surfaces the forces and dynamics of power and identity in which pre-existing knowledge is (re)claimed by sometimes contradictory channels. We have thereby seen that large narrative coalitions form over shared knowledge, which are shaped and held together by linguistic ties, references to over-arching themes (e.g. transhumanism), or a pre-occupation with similar real-world events. It has furthermore been shown how within each of these narrative coalitions, similar forces are at work, leading to similar sources being discussed and pulled on by opposing channels (e.g. the coincidence of Jewish news channels and antisemitic conspiracy channels).

Second, our examination of Telegram channels through the lens of knowledge sources inductively reveals a variety of ideological, epistemological and ontological positions associated with online conspiracism. This includes confluences between science and spiritualism, as well as other idiosyncratic positions that might recontextualize, refute or otherwise subvert established (scientific) evidence. 

Third, our analysis produces further empirical support for the notion of the ``communicability'' of conspiracy theories \cite[p. 13]{milani2020nogo} by demonstrating how narrative coalitions of conspiracism form around shared intertextual links to pre-existing knowledge. These intertextual networks might contribute to the impression that the conspiracy narratives under discussion are both accurate and coherent. It can be further be observed that these conspiracy theories are normalized through their connections - via shared references - with the seemingly mundane and the banal \cite{milani2020nogo}. Specifically, conspiracy channels are embedded within networks that also contain channels devoted to memes, shitposts, popular culture, fandoms, and technology. 

\section{Conclusion and avenues for future research}

By way of conclusion, we highlight the methodological contribution of this study, which may be extended to future work. 

For one thing, the use of transferable methods from the field of bibliometrics proves to be a productive pathway for research into online disinformation and conspiracy theories. This approach, as illustrated here by the use of bibliographic coupling networks, presumes that we can repurpose methods initially designed to study the intellectual structures of established scientific fields to uncover and study similar structures in the realm of (antagonistic) social media. We have referred to such structures as ``narrative coalitions''. In the course of this investigation, it has become clear that methods from bibliometrics can be used as \textit{heuristics} to identify (networks of channels related to) conspiracy theories. The Pushshift dataset under consideration was for instance not originally constructed with the specific aim of studying online conspiracy theories. However, as may follow from the analysis above, tracing bibliographic couplings among channels citing pre-existing knowledge sources proves to be an effective method for identifying meaningful clusters of channels dealing with conspiracy theories (cf. antisemitism) as well as to clusters of channels that recontextualize channels in interesting ways. Our initial experiences here warrant further explorations of how methods that were initially designed to identify deviations from scientific norms in the field might be of interest to identify those phenomena that are of interest to the research in online conspiracy theories and disinformation. 

A second methodological implication of our approach is that it highlights the significance of archival data for the analysis of knowledge and narratives on social media. As access to social media data for research purposes becomes increasingly more restricted, we have demonstrated that valuable insights can be obtained from re-examining historical secondary datasets such as the Pushshift Telegram archive, thus uncovering previously unexplored phenomena and dynamics. Further methodological gains can be expected from cross-referencing such archives with open datasets for bibliometrics research, as we have illustrated here by means of the OpenAlex knowledge graph. By incorporating more fine-grained classifications of knowledge sources (e.g. by domain, by accessibility method (open access or not), or by focusing on scientific research that was later retracted, we expect future research to yield even more elaborate distant readings of how scientific knowledge is refracted and recontextualized in social media texts.

\section{Data and software availability statement}

The Pushshift Telegram dataset analysed in this study is available in its entirety on Zenodo via \url{https://zenodo.org/record/3607497}.
Instructions for downloading data from the OpenAlex knowledge graph are available via \url{https://docs.openalex.org/}.
The Python scripts accompanying this chapter are available on GitHub via \url{https://github.com/willaertt}. 

\section{Ethics and personal data}

No user information was processed for the present research: only data from the messages and channels files were used. For details on how these files were processed and which fields were used in the analysis, we refer to the detailed methodological description in the chapter. 

\section{Funding}

This project has been funded by the European Union’s Horizon Europe programme under grant agreement ID 101094752: Social Media for Democracy (SoMe4Dem) - Understanding the Causal Mechanisms of Digital Citizenship. The funders had no role in study design, data collection and analysis, decision to publish, or preparation of the manuscript.


\end{document}